%% file: chiral.tex
\documentclass[twocolumn,prd,superscriptaddress,longbibliography]{revtex4-2}
\pdfoutput=1
\usepackage{bm}                                                             

\usepackage{graphicx}                                                            
\usepackage{booktabs}                                                        
\usepackage{slashed}                                                             
\usepackage{amssymb,amsmath,amsfonts}                                              
\usepackage{color}                                                                           
\usepackage[utf8]{inputenc}
\usepackage{slashed}
\usepackage{bm}
\usepackage{graphicx}
\usepackage{booktabs}
\usepackage{slashed}
\usepackage{amssymb,amsmath,amsfonts}
\usepackage{color}
\usepackage[utf8]{inputenc}
\usepackage[caption=false]{subfig}

\newcommand{\trace}{{\rm tr}}

\def \beq  {\begin{equation}}
\def \eeq  {\end{equation}}
\def \beqar {\begin{eqnarray}}
\def \eeqar {\end{eqnarray}}
\def\sqr#1#2{{\vcenter{\vbox{\hrule height.#2pt
\hbox{\vrule width.#2pt height#1pt \kern#1pt
\vrule width.#2pt}\hrule height.#2pt}}}}

\definecolor{refcol}{RGB}{178,34,34}
\usepackage{microtype}
\usepackage[colorlinks,linkcolor=refcol,citecolor=refcol,urlcolor=refcol]{hyperref}

\definecolor{red}{rgb}{1,0,0}

\begin{document}

\title{The chiral phase transition and the axial anomaly}


\author{Robert D. Pisarski
}
\affiliation{Department of Physics, Brookhaven National Laboratory, Upton, NY 11973}
\author{Fabian Rennecke
}
\affiliation{Institute for Theoretical Physics, Justus Liebig University Giessen, Heinrich-Buff-Ring 16, 35392 Giessen, Germany}
\affiliation{Helmholtz Research Academy Hesse for FAIR (HFHF), Campus Giessen, Giessen, Germany}
\begin{abstract}
   To date numerical simulations of lattice QCD have not found a chiral phase transition of first order which is expected to occur for sufficiently light pions.  We show how the restoration of an exact global chiral symmetry can strongly decrease the breaking of the approximate, anomalous $U_A(1)$ symmetry.  
   This is testable on the lattice through simulations for one through four flavors. In QCD a small breaking of the $U_A(1)$ symmetry in the chirally symmetric phase generates novel experimental signals.
\end{abstract}

\maketitle

One of the most beautiful phenomena in quantum field theory is the axial anomaly of Adler, Bell, and Jackiw
\cite{Adler:1969gk,Bell:1969ts,Adler:1969er,Bardeen:1969md}.   In four spacetime dimensions
massless fermions are chiral, whose spin is either opposite
or along the direction of motion, and so respectively left or right handed.
For chirally symmetric interactions, as with a gauge field, the current for the total number of fermions, left plus right, is always conserved.  In contrast, the axial current, equal to the difference of the left and right handed currents, is conserved classically
but not quantum mechanically.  Instead, the divergence 
of the axial current is proportional
to the density of the topological charge for the gauge field.

In the vacuum of Quantum ChromoDynamics (QCD), large fluctuations in the topological charge 
explain why the flavor singlet meson, the $\eta '$, is
not a Goldstone boson \cite{tHooft:1976rip,tHooft:1976snw,tHooft:1986ooh}.
It also affects other phenomena, albeit more indirectly
\cite{Veneziano:1989ei,Shore:2007yn, Kharzeev:2015znc, Pisarski:2019upw, Rennecke:2020zgb, Tarasov:2020cwl,Tarasov:2021yll,Giacosa:2017pos,Giacosa:2023fdz}.
The axial anomaly also appears in condensed matter systems \cite{Rylands:2021dds,Fradkin:2023xuy}.

The relationship between the divergence of the axial vector current and topological charge density,
computed at one loop order, is exact \cite{Adler:1969gk,Adler:1969er,Bardeen:1969md}.
Even so, this does not tell one how {\it large} the topologically nontrivial fluctuations are
\cite{Pisarski:1983ms,Nair:2022yqi}.  At zero temperature
they must be large in order to make the $\eta '$ heavy.
In contrast, at high temperature instantons are the dominant
topologically nontrivial fluctuations \cite{Gross:1980br,Boccaletti:2020mxu}.
In this limit the density of instantons can be computed
semiclassically, which implies that 
the magnitude of the topological charge susceptibility vanishes as a high power of the temperature
$T$, as $T \rightarrow \infty$.
Even though it vanishes at high $T$,
it is natural to expect that the density of topologically nontrivial fluctuations is nonzero for any
finite $T$.

This leaves the relationship between the restoration of the exact chiral symmetry, and the approximate, anomalous $U_A(1)$ symmetry, obscure.  Based upon extensive results from numerical simulations
in lattice QCD, in this Letter we outline how the restoration
of an exact chiral symmetry strongly affects the approximate restoration of the anomalous $U_A(1)$ symmetry.  
This can be tested in lattice QCD with different numbers of flavors, especially for a single flavor.  The approximate restoration of the anomalous $U_A(1)$ symmetry has
dramatic implications for the collisions of heavy ions,
and surely implications for condensed matter systems as well.

\section{Effective Lagrangians}
\vspace{-4pt}

We consider QCD-like theories, with a $SU(N_c)$ gauge field coupled to $N_f$ flavors of massless quarks
in the fundamental representation.  
As massless fields, the Lagrangian is invariant under the global chiral rotations
$q_{L,R} \rightarrow {\rm e}^{ i(\theta_V \mp \theta_A)/2} \; U_{L,R} \; q_{L,R}$, where $q_L$ and $q_R$ are left and right handed quarks, and 
$U_L$ and $U_R$ elements of the global symmetry groups $SU_L(N_f)$ and $SU_R(N_f)$, respectively.
There are two $U(1)$ groups, one for quark number, $\theta_V$, and one for axial quark number, $\theta_A$.    

We assume that in vacuum, the exact global chiral symmetry is 
is characterized by an expectation value for a color singlet, spin-zero field $\Phi$,
$\Phi= \overline{q}_L q_R$,
where $\Phi$ transforms under
the fundamental representation of the global symmetry group of
${\cal G}_{\rm cl} = SU_L(N_f) \times SU_R(N_f) \times U_A(1)$
as
$ \Phi \rightarrow {\rm e}^{i \theta_A} \, U_L^\dagger \, \Phi \, U_R.$
As $\Phi$ is invariant under $U_V(1)$, this symmetry can be ignored.

Through the axial anomaly,
the $U_A(1)$ symmetry is violated quantum mechnically by topologically non-trivial fluctuations
such as instantons.  The exact chiral symmetry which remains is just
${\cal G}_{\rm qu} = SU_L(N_f) \times SU_R(N_f)$.

In vacuum, it is expected that chiral symmetry breaks to the maximal diagonal subgroup of $SU_V(N_f)$,
$\langle \Phi^{a b} \rangle = \phi_0 \, \delta^{a b}$,
where $a,b=1\ldots N_f$ are the indices for the $SU_L(N_f)$ and $SU_R(N_f)$ groups.  Phenomenologically, this pattern
certainly occurs in QCD, where $N_c = 3$ and $N_f = 2$ or $3$.
Coleman and Witten proved that it arises in the limit of large $N_c$ and
small $N_f$ \cite{Coleman:1980mx}.

The appropriate effective Lagrangian for chiral symmetry
breaking is well known
\cite{tHooft:1976rip, tHooft:1976snw, tHooft:1980kjq, tHooft:1986ooh, Pisarski:1983ms, Leutwyler:1992yt, Klevansky:1992qe, Jungnickel:1996aa, Creutz:2006ts, Creutz:2009kx, Parganlija:2012fy, Kovacs:2013xca, Grahl:2013pba, Mitter:2013fxa,Grahl:2014fna, Rennecke:2016tkm, Fejos:2015xca, Bass:2018xmz, Horvatic:2018ztu, GomezNicola:2019myi, Pisarski:2019upw, Braun:2020mhk, Fejos:2022mso}.
There are two types of terms which enter.
The first type are terms invariant under ${\cal G}_{\rm cl}$.
Up to terms of sixth order in $\Phi$, these are
\begin{eqnarray}
  \mathcal{L}_{\rm cl} &=&
  \trace(\left|\partial_\mu \Phi\right|^2) + m^2 \, \trace( \Phi^\dagger \Phi)
   \nonumber \\
  &+& \lambda_1\, \big(\trace\, (\Phi^\dagger\Phi) \big)^2\,
  + \lambda_2 \, \trace \big(\Phi^\dagger\Phi\big)^2  \nonumber \\
  &+& \kappa_1\, \big( \trace \, (\Phi^\dagger\Phi)\big)^3
  + \kappa_2 \, \trace \, (\Phi^\dagger \Phi )\; \trace \big(\Phi^\dagger\Phi\big)^2 \nonumber \\
  &&\;\;\;\;\;\;\;\;\;\;+ \kappa_3 \, \trace (\Phi^\dagger \Phi)^3 \; .
  \label{eq:eff_lag_ua1_sym}
\end{eqnarray}
Our trace is normalized, so ${\rm tr}\,  {\bf 1} = 1$.  
For a gauge theory in $3+1$ dimensions,
a phase transition at a nonzero temperature $T$ is characterized by an effective theory in three dimensions.
Couplings to sixth order then represent the relevant operators:
the mass squared, $m^2$; two quartic coupling constants, $\lambda_1$ and $\lambda_2$, with dimensions of mass;
and the six point couplings: $\kappa_1$, $\kappa_2$, and $\kappa_3$, with dimensionless coupling constants.
Terms of eighth and higher order are irrelevant operators, whose coupling constants have negative mass dimension.

The second class of terms are invariant under ${\cal G}_{\rm qu}$ but not $U_A(1)$, and so are generated
by topologically non-trivial fluctuations
\cite{tHooft:1976rip,tHooft:1976snw, tHooft:1986ooh,Pisarski:2019upw,Rennecke:2020zgb}:
\begin{eqnarray}
  \mathcal{L}_{\rm qu} &=& -\xi_1 \, \big(\det\Phi + \det\Phi^\dagger\big)/2 \nonumber \\
  &-& \xi_1^{(1,1)} \, \left( \trace \, \Phi^\dagger \Phi \right) \big(\det\Phi + \det\Phi^\dagger\big)/2 \nonumber \\
  &-& \xi_2 \big[  \big(\det\Phi\big)^2 + \big(\det\Phi^\dagger\big)^2 \big]/2 \; .
\label{eq:eff_lag_ua1_vio}
\end{eqnarray}
For three flavors, these are terms to third, fifth, and sixth order in $\Phi$.

The Atiyah-Singer index theorem relates the change in the axial fermion number
to the topological charge as $n_L - n_R = N_f\, Q$.
An instanton with topological
charge $Q > 0$ has $N_f\, Q$ left handed
zero modes, $q_L$, while for an anti-instanton with $Q < 0$, the quark zero modes are right-handed \cite{Brown:1977bj}.
Thus the first two terms, $\sim \det \Phi$,
arise from instantons with charge one, \cite{tHooft:1976rip,tHooft:1976snw,tHooft:1980kjq,tHooft:1986ooh},
while the last term, $\sim (\det \Phi)^2$, 
is due to instantons with charge two \cite{Pisarski:2019upw,Rennecke:2020zgb}.

We comment that instead of the term $\sim \xi_1$, 
for three flavors Refs.\ \cite{Parganlija:2012fy,Olbrich:2015gln,Divotgey:2016pst,Parganlija:2016yxq} use
$\xi_1 = \xi_1^{(1,1)} = 0$, and just a single anomalous coupling,
$\sim \xi_2' \, [(\det\Phi)^2 - (\det\Phi^\dagger)^2]$.  The operators $\sim\xi_2$ and $\sim \xi_2'$ differ by a term
$\sim \det\Phi^\dagger \det \Phi = \det(\Phi^\dagger \Phi)$.
This operator is invariant under $U_A(1)$, and so for three flavors,
the coupling $\sim \xi_2'$ is equivalent to that $\sim \xi_2$, plus 
a modification of the $U_A(1)$ invariant couplings of
sixth order in Eq. (\ref{eq:eff_lag_ua1_sym}).
A similar relation applies for any number of flavors $\geq 2$.  We prefer to use the coupling $\sim \xi_2$, as that
is uniquely generated by instantons with charge two.

The anomalous couplings in Eq. (\ref{eq:eff_lag_ua1_vio}) are the first terms in an infinite series,
\begin{equation}
  {\cal V}_{\rm qu}(\Phi) = \sum_{i,j,k} \xi^{(j,k)}_i \; (\trace (\Phi^\dagger \Phi)^j)^k
  \left( (\det \Phi)^i + (\det \Phi^\dagger)^i \right)/2 \; .
\end{equation}
Terms with couplings $\sim \xi^{(j,k)}_i$ are generated by fluctuations with topological charge
$|Q| = i$.  For ease of notation, we have denoted $\xi_i^{(0,0)} \equiv \xi_i$.

\section{A conjecture about anomalous couplings}
\vspace{-4pt}
 
At the outset we recognize that especially in vacuum,
the topologically nontrivial configurations are surely
truly quantum objects, and far from any semiclassical
approximation \cite{Nair:2022yqi}.  For ease of
discussion, we refer to the dominant configurations in
vacuum as quantum instantons, and those which dominate
when $T\rightarrow \infty$ as semiclassical instantons.
The contribution of a single semiclassical instanton to
the partition function is
$\sim \exp(- 8 \pi^2 /g^2(T))$, 
so by asymptotic freedom this falls off as a 
high power of the temperature
\footnote{For a $SU(N_c)$ gauge theory
coupled to $N_f$ massless flavors of quarks, the 
topological susceptibility with massive quarks of mass
$m$ falls off as $\sim m^{N_f}/T^{(c-4)}$, $c=(11 N_c- 4N_f C_f)/3$, $C_f=(N^2-1)/(2 N)$.  The factor of $c$ is from the 
leading order coefficient of the $\beta$-function, and
the factor of $-4$ from integration over the instanton scale size. 
 Note that we work in the chiral limit, where $m=0$ and the
 topological susceptibility vanishes as $m^{N_f}$.}. 
Numerical simulations of lattice QCD 
indicate that the topological susceptibility falls off
close to this power down to temperatures of $T_{\rm qu} \sim 300$~MeV
\footnote{The overall magnitude of the topological
susceptibility from a computation of semiclassical
instantons to one loop order is too small by about an
order of magnitude, but surely the two loop
computation is called for.}.  While astonishingly low,
this is still about twice the temperature for
the chiral crossover in QCD, at $T_\chi \sim 156$~MeV
\cite{Kotov:2019dby,Jahn:2021qrp,Borsanyi:2021gqg,Chen:2022fid,Aarts:2023vsf}. Thus we can take $T_{\rm qu}$ as an estimate
of the change from quantum to semiclassical instantons.
\footnote{
It is notable that lattice QCD finds an intermediate region of
$T_\chi < T < T_{\rm qu}$
\cite{Aarts:2023vsf}, where the quantum instantons
have a topological susceptibility different from 
$T > T_{\rm qu}$, but the presence of this intermediate regime does not significantly affect our qualitative analysis.
}.

The essential question is what is the relative 
magnitude of the anomalous coupling constants in vacuum,
and as the temperature increases?
The standard assumption with effective Lagrangians
is that the couplings with the highest mass dimension dominate.
For the $U_A(1)$ symmetric Lagrangian of Eq. (\ref{eq:eff_lag_ua1_sym}), that is the mass squared, followed
by the quartic couplings, {\it etc}.  In the standard Wilsonian
paradigm, this is inescapable, because the only way of differentiating these different operators is through
their mass dimension.

Of course some operators have a larger symmetry than
others: $m^2 \trace(\Phi^\dagger \Phi)$ and
$\lambda_1(\trace(\Phi^\dagger \Phi))^2$ are invariant under
$O(2 N_f^2)$, while the coupling
$\lambda_2 \trace(\Phi^\dagger \Phi)^2$ is only invariant under
$\mathcal{G}_{\rm cl}$.  But this is standard, and doesn't affect the renormalization group flow
\footnote{This includes the possibility that
the symmetry is enlarged at a critical point, as when 
$\lambda_1^{\rm crit} \neq 0$ and $\lambda_2^{\rm crit} = 0$.
But in computing in the renormalization group flow,
both couplings need to be included.}.
The only time that couplings of sixth order need
to be included is at isolated points where both
$\lambda_1$ and $\lambda_2$ vanish; then there is a 
tricritcal point, controlled by the evolution of the
six-point coupling constants, $\kappa_1$, $\kappa_2$, and
$\kappa_3$.

For the anomalous coupling constants, the
operator with the lowest mass dimension is $\xi_1 \det \Phi$.
Thus naively one expects that this operator dominates the
infrared behavior near the chiral phase transition
\cite{Pisarski:1983ms}.

However, there is something special about the anomalous couplings, which is {\it not} true in standard effective theories.  Terms
$\sim \det \Phi$ are due, uniquely, to the zero modes of an
instanton with charge one; those
$\sim (\det \Phi)^2$, to the zero modes of an instanton
with charge two, {\it etc.}
\cite{tHooft:1976rip,tHooft:1976snw,tHooft:1986ooh,Pisarski:2019upw,Rennecke:2020zgb}.

In vacuum, when chiral symmetry breaking
occurs the effective coupling for the {\it first}
anomalous coupling,
$\sim \det \Phi$, is a sum of an infinite number of terms:
\begin{equation}
    \xi_1^{\rm eff}(T) = \sum_{i=1}^\infty\sum_{j,k=0}^{\infty}
    i \, \big(\phi_0(T)\big)^{(i-1)N_f +2 j k} \; \xi_i^{(j,k)}(T) \; .
    \label{eff_xi_one}
    \end{equation}
As indicated, all of the anomalous coupling constants,
the $\xi_i^{(j,k)}$, and the expectation value of the scalar field,
$\phi_0$, are functions of temperature.
At very high temperature, the anomalous coupling constants
$\xi_i^{(j,k)}$ can be computed semiclassically,
and are all nonzero
\cite{tHooft:1976rip,tHooft:1976snw,tHooft:1986ooh,Pisarski:2019upw,Rennecke:2020zgb}. 

We conjecture the following.  In vacuum, the
contribution of $\xi_1(0)$ to the total coupling, 
$\xi_1^{\rm eff}(0)$,
is small.  Instead, terms nominally of higher order
in $\Phi$ in the the effective action
are enhanced by corresponding powers of
the chiral condensate, such as $\sim \phi_0(0)^{N_f} \xi_2(0)$,
$\sim \phi_0(0)^{2 N_f} \xi_3(0)$, {\it etc}.
Our conjecture is that these terms dominate $\xi_1(0)$
numerically.  

In contrast, in
the chirally symmetric phase for $T \geq T_\chi$, the
chiral condensate vanishes, $\phi_0(T) = 0$.
For $T > T_{\rm qu}$, the $\xi_i(T) \sim (\xi_1(T))^i$,
and then $\xi_1(T)$ certainly dominates
over $\xi_{n\geq 2}(T)$.   This is just
because in weak coupling, semiclassical
instantons necessarily form a dilute gas \cite{Pisarski:2019upw,Rennecke:2020zgb}.  

Why should our conjecture be valid?
Consider forming an effective Lagrangian for chiral
symmetry breaking from the underlying gauge theory.  We integrate out quarks and gluons to form an effective theory
for $\Phi$, over some volume $\cal{V}_\chi$.
The essential question is then, what is the distribution
of quantum instantons which contribute in ${\cal V}_\chi$?

If in ${\cal V}_\chi$ quantum
instantons with net charge one dominate,
then so will the operator $\sim \xi_1 \det \Phi$.  If instead
${\cal V}_\chi$ predominately contains 
quantum instantons with net charge
two, then the operator $\sim \xi_2 (\det \Phi)^2$
will be more important.  We suggest, then, that 
in vacuum  
quantum instantons with charge two and greater dominate
$\cal{V}_\chi$.  Of course in all, the topological charge of the vacuum vanishes.  But it need not within a finite volume
$\cal{V}_{\chi}$ 
\footnote{
Lattice QCD finds that in measuring quantities related
to topologically
nontrivial fluctuations, that the autocorrelation
times are much larger than in measuring mass spectra \cite{Aarts:2023vsf}.  Our analysis suggests that
relative to standard hadronic mass scales, that 
quantum instantons are small and/or overlap strongly.
Presumably much finer lattices are required to measure
topological fluctuations.}.

We now discuss the implications of our conjecture,
beginning with the case of three flavors,
which motivated it.

\section{Three flavors}
\vspace{-4pt}

In QCD there is no true phase transition, only
a crossover (albeit with a large increase in the pressure).  If $\xi_1(T_\chi) \neq 0$, however, for three flavors
the operator
$\sim \det \Phi$ is a cubic operator. The presence of a cubic operator implies that the standard effective Lagrangian for a second-order phase transition, with only terms quartic and quadratic in the fields, cannot be reached, and so the transition is of first order. Hence a chiral phase transition of first order {\it must} emerge for sufficiently light pions, $m_\pi < m_\pi^{\rm crit}$ \cite{Pisarski:1983ms}. For simplicity we discuss the case of three degenerate quark flavors.

How large $m_\pi^{\rm crit}$ is depends upon the magnitude
of $\xi_1(T_\chi)$.  
We suggest that in vacuum the $\eta'$ is heavy {\it not} because $\xi_1$ is large, but
because the higher order terms, such as $\xi_2$, $\xi_3$,
{\it etc.}, contribute and overwhelm $\xi_1$.  At the chiral
phase transition, however, $\phi_0 = 0$, and one is left with
just $\xi_1^{\rm eff}(T_\chi) = \xi_1(T_\chi)$.  If $\xi_1(T_\chi)$ is small, then so is $m_\pi^{\rm crit}$.

In mean field theory, it is customary to assume that $\xi_1(T)$ is independent of temperature.
Since the $\eta'$ is so heavy at zero temperature, in vacuum $\xi_1(0)$ must be large, and $m_\pi^{\rm crit}$
should also be large.  In a quark meson model, one finds $m_\pi^{\rm crit} \approx 150$~MeV if the vacuum fluctuations of quarks are ignored \cite{Schaefer:2008hk}, and $m_\pi^{\rm crit} \approx 86$~MeV if they are included \cite{Resch:2017vjs}.
Similarly, using mean field theory in a chiral matrix model yields $m_\pi^{\rm crit} \approx 110$~MeV \cite{skokov_23}. 

Going beyond mean field theory, mesonic fluctuations
can be included by using the functional renormalization group. 
This gives rise to a critical mass which is dramatically smaller but still nonzero, $m_\pi^{\rm crit} \approx 17$~MeV \cite{Resch:2017vjs}. Presumably this occurs because the functional renormalization group is including, at least in part, higher-order anomalous contributions as in Eq.\ (\ref{eff_xi_one}) 
\footnote{We note that the approximation used in Ref.\ \cite{Resch:2017vjs} is known to over-estimate mesonic fluctuations which tend to soften the phase transition \cite{Pawlowski:2014zaa}. But this effect alone cannot explain why $m_\pi^{\rm crit}$ is so small in Ref.\ \cite{Resch:2017vjs}.}.

In contrast, {\it no} simulation of lattice QCD has ever
found evidence of a first order transition.  Instead, they only
place upper bounds on $m_\pi^{\rm crit}$, which are much
smaller than the values in mean field theory.  This includes:
$m_\pi^{\rm crit} < 50$~MeV in Ref.\ \cite{Bazavov:2017xul};
$m_\pi^{\rm crit} < 100$~MeV in Ref.\
\cite{Kuramashi:2020meg}; $m_\pi^{\rm crit} < 90$~MeV in Ref.\ \cite{Dini:2021hug}.
By considering the position of the tricritical point as a function of $N_f$, 
it has been asserted in Ref.\ \cite{Cuteri:2021ikv}
that even for three flavors, the chiral transition is of second order in the chiral limit.

We note that a small value of $\xi_1(0)$ is perfectly
consistent with hadronic phenomenology at zero temperature,
for both hadronic masses and decay widths. In fact, these quantities can be reproduced successfully in low-energy models even with $\xi_2$ as the {\it only} anomalous coupling \cite{Parganlija:2012fy,Kovacs:2013xca,Olbrich:2015gln,Divotgey:2016pst,Parganlija:2016yxq}.
This will be analyzed in greater detail in future work \cite{hungarian}.

While we assume that $\xi_1(T_\chi) \neq 0$, we
stress that we can {\it not} exclude the possibility that $\xi_1(T_\chi)=0$. From the viewpoint of effective Lagrangians, this is most unnatural, as then two parameters ---
$m^2(T)$ and $\xi_1(T)$ --- vanish as one thermodynamic parameter,
the temperature, is varied.

If the result of Ref. \cite{Cuteri:2021ikv} holds
and the chiral transition is of second order, then we 
speculate that not just $\xi_1(T_\chi),$ 
but {\it all} of the anomalous couplings vanish at the critical temperature:
\begin{equation}
    \xi_i^{(j,k)}(T_\chi)=0 \; .
    \label{speculation}
\end{equation}
This implies that the anomalous $U_A(1)$ symmetry is restored at $T_\chi$.  This can only happen precisely at $T_\chi$, since as $T \rightarrow \infty$ semiclassical instantons are present and give $\xi_i^{(j,k)}(T) \neq 0$.  Perhaps Eq. (\ref{speculation})
is a consequence of the 't Hooft anomaly condition \cite{tHooft:1979rat,Fujikawa:2022cee}.

If $\xi_1(T_\chi) = 0$, 
the universality class of a second order chiral
phase transition is that of $\mathcal{G}_{\rm cl}$ symmetry. 
There has been extensive work on this possibility,
including using the $\epsilon$-expansion
\cite{Pisarski:1980ix,Pisarski:1981hir,Pisarski:1983ms},
perturbation theory in three dimensions
\cite{Calabrese:2004uk}, Monte Carlo simulations in three dimensions
\cite{Sorokin:2021jwf,Sorokin:2022zwh}, the functional renormalization group
\cite{Fejos:2022mso}, and the conformal bootstrap
\cite{Nakayama:2014sba, Henriksson:2020fqi,Kousvos:2022ewl}.
As opposed to earlier results 
\cite{Pisarski:1980ix,Pisarski:1981hir,Pisarski:1983ms,Calabrese:2004uk,Sorokin:2021jwf,Sorokin:2022zwh}, recent studies with the functional renormalization group \cite{Fejos:2022mso} and conformal bootstrap \cite{Nakayama:2014sba, Henriksson:2020fqi,Kousvos:2022ewl} find infrared stable fixed points for $\mathcal{G}_{\rm cl}$ in three dimensions.  Thus 
if $\xi_1(T_\chi)=0$, we assume that three massless flavors could have a chiral transition of second order.

This is in accord with recent results using Dyson-Schwinger equations \cite{Bernhardt:2023hpr}, where a second-order chiral transition is found in the chiral limit. In this case scaling analysis shows that the universal physics is described by mean-field behavior without further external input.  A second order transition then arises if $\xi_1(T_\chi)=0$, {\it e.g.}, Refs. \cite{Resch:2017vjs, hungarian}, providing strong indications that this is also true in Ref.\ \cite{Bernhardt:2023hpr}. 

\section{Two and four flavors}
\vspace{-4pt}

For two flavors, the term $\sim \xi_1^{\rm eff}$ is a mass term which splits the $\eta$ meson from
the pions.  The couplings $\sim \xi_1^{(1,1)}$ and $\sim \xi_2$ are of quartic order.  Thus in the chiral
limit, $\xi_1\neq 0$ implies that the $\eta$ meson is massive
at $T_\chi$, and the universality class is that of
$\mathcal{G}_{\rm qu} = SU_L(2) \times SU_R(2) \equiv O(4)$.
Numerical simulations using Wilson fermions by Brandt {\it et al.}~\cite{Brandt:2016daq} find that the mass of the $\eta$ meson is much smaller near $T_\chi$ than at $T=0$, in accord with our conjecture.  If the speculation of
Eq. (\ref{speculation}) holds, then the $\eta$ meson is 
massless at $T_\chi$, and the universality class
is then $O(4) \times O(2)$.

For four flavors, the coupling $\sim \xi_1 \det\Phi$ is of quartic order, and a relevant
quartic coupling, of the same mass dimension as the couplings $\sim \lambda_1$ and $\sim \lambda_2$. 
The critical behavior of $\mathcal{G}_{\rm qu}$ for $N_f = 4$
is unknown.


\section{One flavor}
\vspace{-4pt}

An interesting test of our conjecture is for a single, massless flavor \cite{Leutwyler:1992yt,Creutz:2006ts,Creutz:2009kx}. 
Taking $\Phi = \phi + i \eta$, 
where $\Phi^\dagger \Phi = \phi^2 + \eta^2$, and $(\det\Phi + \det \Phi^\dagger) = 2 \phi$.
Including all couplings to quartic order, the effective Lagrangian is
\begin{eqnarray}
  \mathcal{L}_{\rm qu} &=& (\partial_i \phi)^2 + (\partial_i \eta)^2  + \xi_1 \, \phi \nonumber \\
  &+& m^2 (\phi^2 + \eta^2) + \xi_2 \, (\phi^2 - \eta^2)\nonumber \\
  &+& \xi_3 \, \phi (\phi^2 - \eta^2) + \xi_1^{(1,1)} \, \phi (\phi^2 + \eta^2) \nonumber \\
  &+& \lambda (\phi^2 + \eta^2)^2 \nonumber \\
  &+& \xi_4 \, (\phi^4 - 6 \phi^2 \eta^2 + \eta^4)+ \xi_2^{(1,1)} \, (\phi^4 - \eta^4) \nonumber \; .
\end{eqnarray}
If $\xi_1 \neq 0$ , instantons
directly induce a vacuum expectation value for $\phi$.

If our conjecture is correct, then while there may be no true
chiral phase transition, there could well be a sharp crossover from a low temperature phase, dominated by quantum instantons with large $\xi_1(T)$ and $\phi_0(T)$, to a phase dominated by semiclassical instantons, with small $\xi_1(T)$ and $\phi_0(T)$.  As $T\rightarrow \infty$, $\xi_1(T)$ and $\phi_0(T) \rightarrow 0$.  

If the speculation of Eq. (\ref{speculation}) is true,
only $\lambda(T_\chi) \neq 0$, with
$m^2(T_\chi)$ and {\it all} $\xi_i^{(j,k)}(T_\chi) = 0$.
There is then a chiral phase transition of second order for
an emergent $U_A(1)$ symmetry at $T_\chi$.  This would be most
dramatic.

\section{Implications for QCD}
\label{sec:qcd}
\vspace{-4pt}

We have worked exclusively in the chiral limit.  What are the implications for QCD, where numerical simulations
on the lattice find no true phase transition, but crossover
\cite{Aoki:2006we, HotQCD:2018pds,Borsanyi:2020fev}?

If QCD is close to the chiral limit for three massless flavors, then the restoration of the axial $U_A(1)$ symmetry
at $T_\chi$ surely implies that the {\it approximate} restoration of the axial $U_A(1)$ symmetry is closely
tied to the crossover temperature.

In numerical simulations of lattice QCD, it is common
to measure the violation of the anomalous
$U_A(1)$ symmetry by computing the difference in
the two point functions of pions and $a_0$ mesons
\cite{Dick:2015twa,Tomiya:2016jwr,Ding:2020xlj,Aoki:2020noz,Aoki:2021qws,Kaczmarek:2021ser}.
This is useful for two light flavors, but since for three flavors
a term $\sim \det \Phi$ is cubic in $\Phi$, when
$\phi_0 = 0$ anomalous
terms do not affect mesonic two point functions  \cite{Shuryak:1993ee, Cohen:1996ng, Lee:1996zy, Evans:1996wf}.  
In lattice QCD, at present the situation is unsettled \cite{Kotov:2019dby,Lahiri:2021lrk}: Refs.
\cite{HotQCD:2012vvd,Buchoff:2013nra,Bhattacharya:2014ara,Dick:2015twa,Ding:2020xlj,Kaczmarek:2020sif,Kaczmarek:2021ser,Kaczmarek:2023bxb} find that the anomalous symmetry is not 
even approximately restored by $T_\chi$,
while Refs.
\cite{Brandt:2016daq,Tomiya:2016jwr,Ishikawa:2017nwl,Aoki:2020noz,Aoki:2021qws,Cuteri:2021ikv,Dini:2021hug} find
that it is.  

Our analysis also applies to nonzero quark chemical potential, 
$\mu$.  For a theory at $T\neq 0$, the effective theory is
three dimensional.  If $T \ll \mu$, though, the relevant
effective theory is then in four dimensions.  Assuming that confinement gaps the quarks and gluons,
the effective theory is again that of Eqs.\ (\ref{eq:eff_lag_ua1_sym}) and (\ref{eq:eff_lag_ua1_vio}).
While the mass dimensions of the coupling constants change,
the conclusion remains that if $\xi_1(T_\chi) \neq 0$, the chiral phase
transition is of first order in the chiral 
limit.
 
Our analysis predicts that
the breaking of the anomalous $U_A(1)$ symmetry is uniformly
{\it small} in a chirally symmetric regime.  The $\eta '$
meson, which is heavy is vacuum, must become light.

There is an interesting possibility which arises.
Like the $U_A(1)$ invariant coupling constants,
the anomalous coupling constants are all
functions of {\it both} temperature and chemical
potential, $\xi_i^{(j,k)}(T,\mu)$. Analogous to
the critical endpoint, where for two light flavors the 
$O(4)$ invariant quartic coupling constant vanishes, 
$\lambda(T^{\rm cr},\mu^{\rm cr})=0$
\cite{Asakawa:1989bq,Stephanov:1998dy,Stephanov:1999zu,Bzdak:2019pkr}, since we have two thermodynamic parameters to vary,
it is possible that there is a {\it single} point
in the phase diagram where $\xi_1(T^{A},\mu^{A})=0$.
About this point,
instead of $SU_V(3)$ flavor eigenstates, the $\pi^0$, $\eta$, and $\eta'$ are eigenstates of flavor, and there is a large
violation of isospin \cite{Pisarski:1983ms}. 
It is very intriguing that such a violation has been reported by the NA61/SHINE collaboration recently
\cite{NA61SHINE:2023azp,Brylinski:2023nrb}.

If $\xi_1(T,\mu)$ vanishes at a point in the plane
of $T$ and $\mu$, then
perhaps there is a region where $\xi_1(T,\mu)$ if of 
{\it opposite}
sign to that in the vacuum.  If chiral symmetry
is broken, then instead of the $\sigma$ meson condensing,
the $\eta '$ does.  This implies that $CP$ symmetry
is spontaneously broken by an $\eta '$ condensate.

Other signals which have been suggested include:
Hanbury-Brown-Twiss correlations \cite{Vance:1998wd,Csorgo:2009pa,Vertesi:2009wf}, 
possibly confirmed by the PHENIX experiment \cite{PHENIX:2017ino}, 
and an excess of soft dileptons \cite{Vargyas:2012ci}.
The HADES experiment finds that the $\eta$ meson is about twice
as abundant as expected from a statistical distribution \cite{HADES:2015oef,HADES:2019auv}.
Certainly when the $\eta '$ meson becomes light, so does
the $\eta$ meson \footnote{We note that other effects can also contribute to the $\eta$ meson abundance, cf., e.g., Ref.\ \cite{Larionov:2021ycq}.}.

Besides the other implications of our results, it is also
natural to wonder how the suppression of topologically nontrivial
fluctuations in a chirally symmetric phase affects baryogenesis
in the early universe \cite{Bodeker:2020ghk}.

\acknowledgments
R.D.P. is supported by the U.S. Department of Energy under contract DE-SC0012704, and thanks the Alexander V. Humboldt
Foundation for their support.
F.R. is supported by the Deutsche Forschungsgemeinschaft (DFG, German Research Foundation) through the Collaborative
Research Center TransRegio CRC-TR 211 “Strong- interaction matter under extreme conditions” – project number
315477589 – TRR 211.
We thank J.\ Bernhardt, M.\ Creutz, T.\ Csorgo, H.\ Davoudiasl, C.\ Fischer, T.\ Galatyuk, F.\ Giacosa, K.\ Intriligator, D.\ Kaplan, S.\ Kousvos, G.\ Kovacs, P.\ Kovacs,
S.\ Mukherjee, V.\ P.\ Nair, P.\ Petreczky, O.\ Philipsen, A.\ Sorokin, R. Szafron, and A.\ Stergiou for discussions.

\input{chiral.bbl}

\end{document}

%% file: chiral.bbl
%